\documentclass[useAMS]{mn2e}
\usepackage{rotating}
\usepackage{mncite}

\def\approxgt{\ifmmode \rlap{$>$}{}_{{}_{{}_{\textstyle\sim}}} \else%
$\rlap{$>$}{}_{{}_{{}_{\textstyle\sim}}}$\fi} 
\def\approxlt{\ifmmode \rlap{$<$}{}_{{}_{{}_{\textstyle\sim}}} \else%
$\rlap{$<$}{}_{{}_{{}_{\textstyle\sim}}}$\fi}

\LARGE \normalsize \title[]{The faint neutron star soft X--ray
transient SAX~J1810.8--2609 in quiescence}

\author[P.G. Jonker et al.]  {P.G. Jonker$^1$\thanks{email :
    peterj@ast.cam.ac.uk}, R. Wijnands$^2$, M. van der Klis$^3$ \newauthor \\
  $^1$Institute of Astronomy, Madingley Road, CB3 0HA, Cambridge, UK\\
  $^2$School of Physics and Astronomy, University of St Andrews, North Haugh, St Andrews, KY16 9SS, Fife, Scotland, UK\\
  $^3$Astronomical Institute ``Anton Pannekoek'',
  University of Amsterdam, Kruislaan 403, 1098 SJ Amsterdam, The Netherlands\\
}

\begin{document}

\maketitle

\begin{abstract}
\noindent 
We present the analysis of a 35 ksec long {\it Chandra} observation of
the neutron star soft X--ray transient (SXT) SAX~J1810.8--2609. We
detect three sources in the field of view. The position of one of them
is consistent with the location of the ROSAT error circle of
SAX~J1810.8--2609. The accurate {\it Chandra} position of that source
coincides with the position of the proposed optical counterpart,
strengthening the identification as the counterpart. We detected the
neutron star SXT system in quiescence at an unabsorbed luminosity of
$\sim1\times10^{32}$ erg s$^{-1}$ (assuming a distance of 4.9 kpc).
This luminosity is at the low--end of quiescent luminosities found in
other neutron star SXTs. This renders support to the existence of a
group of faint soft X--ray transients of which the accreting
millisecond X--ray pulsar SAX~J1808.4--3658 is the most prominent
member. The quiescent spectrum of SAX~J1810.8--2609 is well--fit with
an absorbed power law with photon index of $3.3\pm0.5$.  With a value
of 3.3$\times10^{21}$ cm$^{-2}$ the Galactic absorption is consistent
with the value derived in outburst. Since the spectra of quiescent
neutron star SXTs are often fit with an absorbed blackbody or neutron
star atmosphere plus power--law model we also fitted the spectrum
using those fit functions. Both models provide a good fit to the data.
If cooling of the neutron star core and/or crust is responsible for
the soft part of the spectrum the time averaged mass accretion rate
must have been very low ($\sim$5.7$\times10^{-13}$ M$_\odot$
yr$^{-1}$; assuming standard core cooling only) or the neutron star
must be massive. We also discuss the possibility that the thermal
spectral component in neutron stars in quiescence is produced by
residual accretion.

\end{abstract}


\section{Introduction}
\label{intro}

Low--mass X--ray binaries are binary systems in which a
$\approxlt$1\,$M_{\odot}$ star transfers matter to a neutron star or a
black hole. A large fraction of the low--mass X--ray binaries is
transient; these systems form the so called soft X--ray transients
(SXTs). The characterising property of SXTs is that the accretion rate
drops several orders of magnitude when the source returns to
quiescence (\pcite{1984heta.conf...49V}).

SXTs have been and are being studied extensively with previous and
present X--ray satellites; for an overview of SXTs see
\scite{1997ApJ...491..312C}. Using data obtained with the Beppo{\it
  SAX} satellite \scite{1999ApL&C..38..297H} and
\scite{2001ESASP.459..463I} suggested that $\sim$10 bursting neutron
stars form a separate class of {\it faint} SXTs; the outburst peak
luminosity is low (typically $\sim10^{36.5}$ erg s$^{-1}$). Later,
\scite{2000MNRAS.315L..33K} argued that a class of faint SXT could be
explained evolutionarily. \scite{2000MNRAS.315L..33K} argues that
faint SXTs should have evolved beyond the period minimum of $\sim$80
minutes to orbital periods of 80--120 minutes.

Using the {\it Chandra} and {\it XMM/Newton} satellites it is possible
to study these systems even when they are in quiescence
(cf.~\pcite{2001ApJ...560L.159W}; \pcite{2002ApJ...580..413R};
\pcite{2002ApJ...575L..15C}). A large fraction of the neutron star SXT
quiescent X--ray luminosity is often ascribed to cooling of the hot
neutron star core and/or crust (cf.~\pcite{1987A&A...182...47V};
\pcite{1998ApJ...504L..95B}).  Assuming that the thermal spectral
component is coming from the neutron star surface, in principle
neutron star parameters such as the mass and radius can be derived
when using a neutron star atmosphere model to fit the spectrum
(\pcite{2001ApJ...551..921R}; \pcite{2001ApJ...559.1054R};
\pcite{2003ApJ...588..452H}).  The quiescent luminosity of the neutron
star SXT and accreting millisecond X--ray pulsar SAX~J1808.4--3658 was
found to be very low (L$_X=5\times10^{31}$ erg s$^{-1}$;
\pcite{2002ApJ...575L..15C}). This is nearly as low as the
luminosities found for several quiescent BHC SXTs with a short orbital
period (\pcite{2002ApJ...570..277K}).  Such a low neutron star
luminosity could result if the neutron star core and crust are
relatively cool (\pcite{2001ApJ...548L.175C}). This in turn could hint
at a massive neutron star (M$_{NS}>1.7$ M$_\odot$;
\pcite{2001ApJ...548L.175C}), or a very low time--averaged mass
accretion rate (\pcite{1998ApJ...504L..95B}).

Several other emission mechanisms have been proposed to explain the
quiescent luminosity of SXTs as well. First, in quiescence mass
accretion may be ongoing at a low level (\pcite{1995ApJ...439..849Z},
possibly via an advection dominated accretion flow,
e.g.~\pcite{1999ApJ...520..276M}). Second, shocks formed by the
propeller mechanism preventing accretion onto the neutron star may
produce X--rays (\pcite{1975A&A....39..185I};
\pcite{1986ApJ...308..669S}). Third, the switch--on of a radio pulsar
mechanism may produce the observed X--rays
(\pcite{2000ApJ...541..849C}).  Finally, leaking of matter through the
magnetospere especially at high latitudes in an ADAF--like spherical
flow may produce X--rays (e.g.~\pcite{1998ApJ...494L..71Z}).

The quiescent source luminosity seems to be a characterising
difference between the systems containing a neutron star and those
containing a black hole candidate (BHC) compact object.  When plotted
versus the binary orbital period the neutron star SXTs seem to be
systematically brighter by a factor of $\sim$10 than the BHC SXTs
(cf.~\pcite{2001ApJ...553L..47G}). A possible and exciting explanation
is that in the BHC SXTs energy is advected across the event horizon
and not thermalised and radiated away like in neutron star SXTs
(\pcite{1997ApJ...482..448N}; \pcite{2001ApJ...553L..47G}).  However,
\scite{2003MNRAS.343L..99F} suggested that the difference between the
quiescent X--ray luminosities from neutron stars and BHC SXTs could
also be explained by taking into account the jet power.

SAX~J1810.8--2609 was discovered with the Wide Field Cameras on board
the Beppo{\it SAX} satellite (\pcite{1998IAUC.6838....1U}). A type~I
X--ray burst was discovered establishing that the compact object is a
neutron star (\pcite{1998IAUC.6838....1U};
\pcite{1999ApL&C..38..133C}). The burst showed evidence for radius
expansion which led to a distance estimate of $\sim4.9$ kpc
(\pcite{2000ApJ...536..891N}). \scite{1999MNRAS.308L..17G} obtained
ROSAT HRI follow--up observations and reported a candidate optical and
near--infrared counterpart of SAX~J1810.8--2609.

In this Paper we present the analysis of our 35 ksec {\it Chandra}
observation of the faint neutron star SXT SAX~J1810.8--2609 while the
source is in quiescence.

\section{Observations and analysis}
We have observed the neutron star SAX~J1810.8--2609 using the ACIS--S3
CCD (windowed to 1/8 of its original size to reduce possible pile--up
if the source was found to be relatively bright) on board the {\it
  Chandra} satellite (\pcite{1988SSRv...47...47W}) for $\sim$35 ksec
on August 16, 2003.  The observation started at 21:20:35 Terrestrial
Time.

The X--ray data were processed by the {\it Chandra} X--ray Centre;
events with ASCA grades of 1, 5, 7, cosmic rays, hot pixels, and
events close to CCD node boundaries were rejected. We used the
standard {\it CIAO} software to reduce the data (version 3.0.1 and CALDB
version 2.23). We searched the data for periods of enhanced background
radiation but none was present. Hence, all the data were used in our
analysis.

We detect three sources using the {\it CIAO} tool {\sl celldetect} and
derive the following coordinates for them: \newline R.A.=18h10m44.47s,
Decl.=-26$^\circ$09'01.2" \newline
\noindent
R.A.=18h10m43.49s, Decl.=-26$^\circ$10'44.0"  \newline
\noindent
R.A.=18h10m42.03s, Decl.=-26$^\circ$11'02.9" \newline
\noindent
All positions have a typical error of 0.6" and use equinox J2000.0.
Only the coordinates of the first source are consistent with the ROSAT
coordinates obtained for SAX~J1810.8--2609
(\pcite{1999MNRAS.308L..17G}). The coordinates of the proposed optical
counterpart of SAX~J1810.8--2609 (R.A.=18h10m44.4s,
Decl.=-26$^\circ$09'00", uncertainty 1 arcsec;
\pcite{1999MNRAS.308L..17G}) are consistent with those of the first
source. We conclude that we have detected SAX~J1810.8--2609 in
quiescence and that the proposed counterpart is very likely correct.
We designate the other two sources CXOU~J181043.5--261044 and
CXOU~J181042.0--261103, respectively. The spectrum of both sources can
be fit by an absorbed power law spectrum.  No obvious counterparts
were found in the second digital sky survey red nor 2MASS K--band
catalogue. Both sources are likely background AGNs.

SAX~J1810.8--2609 is detected at a count rate of
(4.0$\pm$0.4)$\times10^{-3}$ counts s$^{-1}$. The source spectrum was
extracted from a region with a 5 arc second radius centred on the
source whereas the background spectrum was extracted from an annulus
centred on the source with an inner and outer radius of 10 and 20
arc seconds, respectively. We extracted the spectrum with at least 10
counts per bin. Because of this relatively low number of counts we
checked the results obtained using the $\chi^2$ fitting method against
the results obtained using the CASH statistic method
(\pcite{1979ApJ...228..939C}); the fit--results were consistent within
the 1$\sigma$ error bars. We only include energies above 0.3 and below
10 keV in our spectral analysis since the ACIS timed exposure mode
spectral response is not well calibrated outside that range. We fit
the spectra using the {\sc XSPEC} package (version 11.2.0bp;
\pcite{1996adass...5...17A}). We included the multiplicative {\sl
  ACISABS} model in our spectral fitting to account for the additional
absorption due to contamination by the optical blocking filters in our
spectral fits (\pcite{astroph.marshall}).

\begin{figure*}
  \includegraphics[angle=-90,width=15cm]{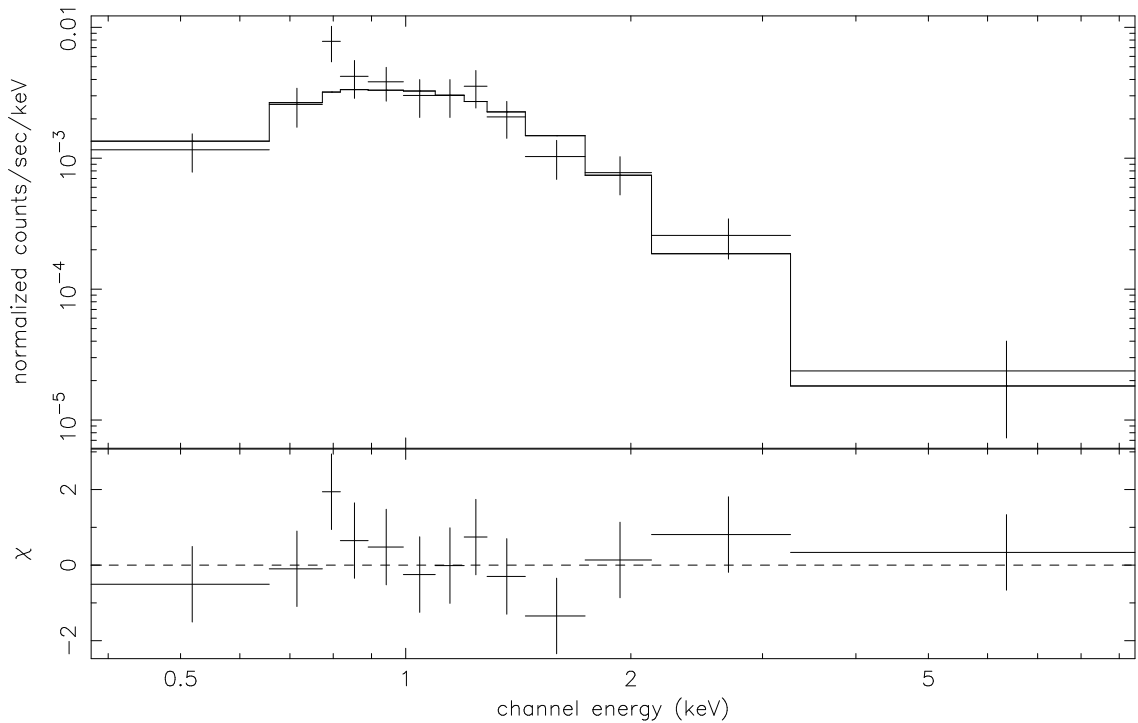}
\caption{The quiescent X--ray spectrum (0.3--10 keV) of
  SAX~J1810.8--2609 fitted with an absorbed power law
  model. The residuals (data minus model) are plotted
  in the bottom panel. }
\label{pl}
\end{figure*}

We tried to fit the spectrum with several absorbed single--model
fit--functions (i.e.~an absorbed blackbody, power law, neutron star
atmosphere [NSA, \pcite{1991MNRAS.253..193P};
\pcite{1996A&A...315..141Z}], and bremsstrahlung model) but only the
absorbed power--law model provides a good fit ($\chi_{red}^2=0.8$ for
10 degrees of freedom (d.o.f.); see Figure~\ref{pl}). The
$\chi_{red}^2$ for the NSA, blackbody and bremsstrahlung model were
1.6 for 11 d.o.f., 1.9 for 11 d.o.f., and 1.4 for 11 d.o.f.,
respectively and in all cases large residuals at energies above 2 keV
were apparent. However, with a value of 3.3$\pm$0.5, the best--fit
power--law photon index is high. The derived interstellar absorption
of 3.3$\pm0.8\times10^{21}$ cm$^{-2}$ is consistent with the value
derived by \scite{2000ApJ...536..891N} using Beppo{\it SAX} outburst
data of the source and also with that derived by
\scite{1990ARA&A..28..215D} (N$_H\sim4\times10^{21}$ cm$^{-2}$). We
present the best--fit parameters for the single absorbed power--law
model in Table~\ref{fitpars}. 

The high best--fit power--law index when fitting a single absorbed
power law indicates that the spectrum is rather soft. Since two
component models are often used for neutron star SXTs in quiescence we
also fitted the spectrum using an absorbed NSA plus power--law model.
The Galactic absorption, the NSA normalisation, the mass, and the
radius of the neutron star were held fixed during the fit at a value
of $3.3\times10^{21}$ cm$^{-2}$, $\frac{1}{D^2}=4.16\times10^{-8}$
pc$^{-2}$ (for the distance, D in pc, we took the value derived by
\pcite{2000ApJ...536..891N} 4.9$\times10^3$ pc), 1.4 M$_\odot$, and 10
km, respectively.  The fit was good with a $\chi_{red}^2$ of 0.65 for
10 degrees of freedom. The best--fit parameters are given on the
second line of Table~\ref{fitpars}.  The absorbed 0.5--10 keV source
flux was $\sim$2.1$\times10^{-14}$ erg cm$^{-2}$ s$^{-1}$, whereas the
unabsorbed flux was $\sim$3.8$\times10^{-14}$ erg cm$^{-2}$ s$^{-1}$.
The neutron star atmosphere and power law models each contribute about
50 per cent of the unabsorbed flux. Finally, we fitted an absorbed
blackbody plus power--law model to the spectrum. With a $\chi_{red}^2$
of 0.62 for 9 degrees of freedom the fit was good. The blackbody
contributed about 40 per cent to the total unabsorbed flux and the
power law about 60 per cent. The best--fit parameters are given on the
third line in Table~\ref{fitpars}.

\begin{table*}
\caption{Best fit parameters of the quiescent spectrum of SAX~J1810.8--2609. 
All quoted errors are at the 68 per cent confidence level. The second line
displays the best--fit temperature for the neutron star atmosphere (NSA; at 
the surface) plus power--law model, whereas the third line shows those of the 
blackbody (BB) plus power--law (PL) model.}

\label{fitpars}
\begin{center}

\begin{tabular}{lccccc}
\hline 
Model & Power law & &  BB radius & Temperature & Reduced \\ 
 & Index & 10$^{-6}$ photons keV$^{-1}$ cm$^{-2}$ s$^{-1, a}$ & in km, assuming D=4.9 kpc& keV & $\chi^2/$d.o.f.\\
\hline 
\hline
PL & 3.3$\pm$0.5 & 16$\pm$6 & --- &  --- &  0.80/10 \\ 
NSA + PL & 2.3$\pm0.8$ & 6$^{+8}_{-4}$ &  --- & 0.072$\pm0.004$&  0.65/10 \\
BB + PL& 2.5$\pm$0.4 & 8$\pm5$ & 40$_{-30}^{+60}$ & 0.14$\pm0.02$ &  0.62/9\\
\end{tabular}
\end{center}

{\footnotesize $^a$ Power
law normalisation at 1 keV.}\newline 

\end{table*}

\section{Discussion}

We detected X--ray emission from three sources using data obtained
with the {\it Chandra} satellite. The coordinates of only one of the
sources are consistent with the ROSAT position of the neutron star SXT
SAX~J1810.8--2609. We conclude that we detected SAX~J1810.8--2609 in
quiescence. The source position is also fully consistent with the
proposed optical--infrared counterpart of SAX~J1810.8--2609
(\pcite{1999MNRAS.308L..17G}) strengthening the identification. The
source spectrum is well described with an absorbed neutron star
atmosphere, or blackbody, plus power--law model. With a value of 0.14 or
0.07 keV, the temperature of the neutron star as fitted with the
blackbody and the NSA model, respectively, is low compared with many
other neutron star SXTs in quiescence where values around 0.2--0.4 keV
for the blackbody or 0.1--0.2 keV for the NSA model are usual
(cf.~\pcite{1996PASJ...48..257A}; \pcite{2001ApJ...559.1054R}, but see
also \pcite{2001ApJ...551..921R}).  A single absorbed power law also
gave a good fit. However, with a power--law index of $\sim$3.3 it was
rather soft.  The unabsorbed 0.5--10 keV source flux and luminosity
are $\sim4\times10^{-14}$ erg cm$^{-2}$ s$^{-1}$ and
$\sim1\times10^{32}$ erg s$^{-1}$, respectively.  In calculating the
luminosity we used the source distance of 4.9 kpc as determined by
\scite{2000ApJ...536..891N}. The outburst peak luminosity of
SAX~J1810.8--2609 was $\sim2\times10^{36}$ erg s$^{-1}$
(\pcite{2000ApJ...536..891N}).

The outburst and quiescent luminosity are at the low end of typical
outburst and quiescent luminosities of neutron star SXTs of
$\sim2\times10^{38}$ erg s$^{-1}$ and several times $\sim10^{32-33}$
erg s$^{-1}$, respectively. Furthermore, they are close to those of
the accreting millisecond X--ray pulsar SAX~J1808.4--3658
($\sim2\times10^{36}$ erg s$^{-1}$, and $\sim5\times10^{31}$ erg
s$^{-1}$, for a distance of 2.5 kpc; \pcite{1998A&A...331L..25I};
\pcite{1998Natur.394..344W}; \pcite{2001A&A...372..916I};
\pcite{2002ApJ...575L..15C}). Both systems were proposed to be members
of the class of recently recognised sub--group of {\it faint} neutron
star SXTs (\pcite{1999ApL&C..38..297H}; \pcite{2001ESASP.459..463I}).
Obviously, if one would underestimate the source distance of the faint
systems by a factor of a few one could derive anomalously low values
for the luminosity. However, in both these neutron stars, radius
expansion bursts have been found (see references above). As was shown
by \scite{2003A&A...399..663K} using a sample of neutron stars in
Globular Clusters exhibiting radius expansion bursts, the typical
error in the distance estimate as derived from the radius expansion
burst is 15 per cent. This is insufficient to explain the difference
in outburst and quiescent luminosity between the faint and 'normal'
neutron star SXTs. The low luminosity of SAX~J1810.8--2609 in
quiescence hence strengthens the case for the existence of the class
of faint (neutron star) SXTs.

The quiescent luminosity of neutron star SXTs has been ascribed to
thermal cooling of the hot neutron star crust/core
(\pcite{1998ApJ...504L..95B}). However, this interpretation is not
without problems. As was noted by several authors (among which
\pcite{1998ApJ...504L..95B} themselves, see also
\pcite{2000ApJ...541..849C}), thermal cooling cannot explain the
power--law spectral component. As mentioned in the introduction, a low
quiescent luminosity could arise if the neutron star is more massive
than $\sim1.7$ M$_\odot$ since in that case enhanced core cooling due
to the direct URCA process could take place
(\pcite{2001ApJ...548L.175C}). However, the neutron star core could
also be cold if the time--averaged mass accretion rate onto the
neutron star is very low (\pcite{1998ApJ...504L..95B}). In case of
SAX~J1810.8--2609 the time average mass accretion rate necessary to
explain the flux coming from the thermal spectral component would be
5.7$\times10^{-13}$ M$_\odot$ yr$^{-1}$ assuming that the amount of
heat deposited in the crust per accreted nucleon is 1.45 MeV (assuming
that only standard core cooling processes are at work; see
\pcite{1998ApJ...504L..95B}).

Initially, the relatively large variability found in the temperature
associated with the soft component in Cen~X--4 and Aql~X--1 was
regarded as a problem for the cooling neutron star model
(\pcite{1999ApJ...514..945R}; \pcite{2001ApJ...559.1054R}) and only if
the neutron stars are massive could the variability be explained
(\pcite{2001MNRAS.325.1157U}).  However, \scite{2002ApJ...574..920B}
found that since the residual hydrogen and helium mass in the neutron
star atmosphere varies from one outburst to the next the thermal
quiescent flux can vary between outbursts by a factor of 2--3.
However, \scite{2002ApJ...577..346R} found that the quiescent thermal
flux of Aql~X--1 after an outburst first decreased by up to 50 per
cent and then increased again by 35 per cent on timescales of a month.
It also varied by nearly 40 per cent on short time scales (ksec).
\scite{2002ApJ...577..346R} concluded that accretion ongoing at a low
level is the most likely explanation (however, it is unclear why the
the luminosity agrees well with the predictions from deep crustal
heating. See also \pcite{2003ApJ...597..474C} for an alternative
explanation).  \scite{1987A&A...182...47V} also considered the
possibility that mass accretion is ongoing at a low rate when the
systems are in quiescence.  \scite{1995ApJ...439..849Z} showed that
emergent spectra of neutron stars accreting at a low rate would
resemble a thermal spectrum albeit overall hardened.

We investigated the observational findings of neutron star SXTs in
quiescence.  There is evidence that in sources which have a low
quiescent source luminosity the contribution of the power--law
component to the spectrum is large compared with more luminous neutron
star sources in quiescence. The source with the lowest luminosity,
SAX~J1808.4--3658 has a spectrum which is well--fitted with just a
power--law (\pcite{2002ApJ...575L..15C}). Other sources like Cen~X--4
and SAX~J1810.8--2609 with luminosity, L,
10$^{32}\approxlt$L$\approxlt5\times10^{32}$ erg s$^{-1}$ show a
power--law component which contributes up to about 50 per cent of the
source luminosity (cf.~\pcite{1996PASJ...48..257A};
\pcite{2001ApJ...551..921R}; this work). Sources with luminosity near
10$^{33}$ erg s$^{-1}$ like X~5 and X~7 in the Globular Cluster
47~Tucanae (\pcite{2003ApJ...588..452H}) are well--fit without the
power--law component. Most quiescent neutron star SXTs in Globular
Clusters fit this trend (the addition of a power--law component to the
fit was never necessary to obtain a good fit,
\pcite{2003ApJSubmittedHetal}). However, the neutron star SXT in
Terzan 5 and Aql~X--1 complicate the picture.  Their quiescent
spectrum is dominated by or in the case of Aql~X--1 has a power--law
component even though the quiescence luminosity is high
$\sim$2--4$\times$10$^{33}$ erg s$^{-1}$ (for Terzan 5 see
\pcite{wijnsubter5}; for Aql~X--1 see
e.g.~\pcite{2001ApJ...559.1054R}; \pcite{2002ApJ...577..346R}).
Perhaps the power--law spectral component at source luminosities
higher than 10$^{33}$ erg s$^{-1}$ has a different origin than those
at lower source luminosities. 

Finally, as was noted before (cf.~\pcite{2003ApJ...594..952W}), we
like to point out that assuming the observed black body radiation is
caused by residual accretion and not a cooling neutron star crust the
power--law spectral component at the lowest source luminosities
(i.e.~those found in SAX~J1808.4--3658) could be similar in nature to
the one found for black hole candidate SXTs in quiescence. This could
tie in with the neutron stars entering a jet--dominated state when the
accreting rate drops (\pcite{2003MNRAS.343L..99F}).

\section*{Acknowledgments} 
\noindent 
PGJ is supported by EC Marie Curie Fellowship HPMF--CT--2001--01308.
MK is supported in part by a Netherlands Organisation for Scientific
Research (NWO) grant. We would like to thank the referee for his/her
comments which improved the paper.

\end{document}